# Using Association Rules for Better Treatment of Missing Values

SHARIQ BASHIR, SAAD RAZZAQ, UMER MAQBOOL, SONYA TAHIR,
A. RAUF BAIG
Department of Computer Science (Machine Intelligence Group)
National University of Computer and Emerging Sciences
A.K.Brohi Road, H-11/4, Islamabad.
Pakistan
shariq.bashir@nu.edu.pk, rauf.baig@nu.edu.pk

*Abstract:* - The quality of training data for knowledge discovery in databases (KDD) and data mining depends upon many factors, but handling missing values is considered to be a crucial factor in overall data quality. Today real world datasets contains missing values due to human, operational error, hardware malfunctioning and many other factors. The quality of knowledge extracted, learning and decision problems depend directly upon the quality of training data. By considering the importance of handling missing values in KDD and data mining tasks, in this paper we propose a novel Hybrid Missing values Imputation Technique (HMiT) using association rules mining and hybrid combination of k-nearest neighbor approach. To check the effectiveness of our HMiT missing values imputation technique, we also perform detail experimental results on real world datasets. Our results suggest that the HMiT technique is not only better in term of accuracy but it also take less processing time as compared to current best missing values imputation technique based on k-nearest neighbor approach, which shows the effectiveness of our missing values imputation technique.

## 1 Introduction

The goal of knowledge discovery in databases (KDD) and data mining algorithms is to form generalizations, from a set of training observations, and to construct learning models such that the classification accuracy on previously unobserved observations are maximized. For all kinds of learning algorithms, the maximum accuracy is usually determined by two important factors: (a) the quality of the training data, and (b) the inductive bias of the learning algorithm.

The quality of training data depends upon many factors [1], but handling missing values is considered to be a crucial factor in overall data quality. For many real-world applications of KDD and data mining, even when there are huge amounts of data, the subset of cases with complete data may be relatively small. Training as well as testing samples have missing values. Missing data may be due to different reasons such as refusal to responds, faulty data collection instruments, data entry problems and data transmission problems. Missing data is a problem that continues to plague data analysis methods. Even as analysis methods gain sophistication, we continue to encounter missing values in fields, especially in databases with a large number of fields. The absence of information is rarely beneficial. All things being equal, more data is almost always better. Therefore, we should think carefully about how we handle the thorny issue of missing data.

Due to the frequent occurrence of missing values in training observations, imputation or prediction of the missing data has always remained at the center of attention of KDD and data mining research community [13]. Imputation is a term that denotes the procedure to replace the missing values by considering the relationships present in the observations. One main advantage of imputing missing values during preprocessing step is that, the missing data treatment is independent of the learning algorithm. In [6, 7] imputing missing values using prediction model is proposed. To impute missing values of attribute *X*, first of all a prediction model is constructed by considering the attribute *X* as a class label and other attributes as input to prediction model. Once a prediction model is constructed, then it is utilized for predicting missing values of attribute *X*. The main advantages of imputing missing values using this approach are that, this method is very useful when strong attribute relationship exists in the training data, and secondly it is a very fast and efficient method as compared to k-nearest neighbor approach. The imputation processing time depends only on the construction of prediction model, once a prediction model is constructed, then the missing values are imputed in a constant time. On the other hand, the main drawbacks of this approach are that, if there is no relationship exists among one or more attributes in the dataset and the attributes with missing data, then the prediction model will not be suitable to estimate the missing value. The second drawback of this approach is that, the predicted values are usually more accurate than the true values.

In [6] and [7] another important imputation technique based on k-nearest neighbor is used to impute missing values for both discrete and continuous value attributes. It uses majority voting for categorical attributes and mean value for continuous value attributes. The main advantage of this technique it that, it does not require any predictive model for missing values imputation of an attribute. The major drawbacks of this approach are the choice of using exact distance function, considering all attributes when attempting to retrieve the similar type of examples, and searching through all the dataset for finding the same type of instances, require a large processing time for missing values imputation in preprocessing step.

To overcome the limitations of missing values imputation using prediction model and decrease the processing time of missing values treatment in preprocessing step, we propose a new missing values imputation technique (*HMiT*) using association rules mining [2, 3, 9] and hybrid combination of k-nearest neighbor approach [6, 7]. Association rule mining is one of the major techniques of data mining and it is perhaps the most common form of local-pattern discovery in unsupervised learning systems. Before this, different forms of association rules have been successfully applied in the classification [9, 12], sequential patterns [4] and fault-tolerant patterns [14]. In [9, 12] different extensive experimental results on real world datasets show that classification using association rule mining is more accurate than neural network, Bayesian classification and decision tress.

In *HMiT*, the missing values are imputed using association rules by comparing the known attribute values of missing observations and the antecedent part of association rules. In the case, when there is no rule present or fired (i.e. no attributes relationship exists in the training data) against the missing value of an observation, then the missing value is imputed using k-nearest neighbor approach. Our experimental results suggest that our imputation technique not only increases the accuracy of missing values imputation, but it also sufficiently decrease the processing time of preprocessing step.

## 2 Hybrid Missing Values Imputation Technique (HMiT)

Let $I = \{i_1 ... i_n\}$ be the set of n distinct items. Let *TDB* represents the training dataset, where each record *t* has a unique identifier in the *TDB*, and contains set of items such that $t_{items} \subset I$. An association rule is an expression $A \rightarrow B$, where *A* and *B* are items $A \subset I$, $B \subset I$, and $A \cap B = \phi$. Here, *A* is called antecedent of rule and *B* is called consequent of rule.

The main technique of *HMiT* is divided into two phases- (a) firstly the missing values are impute on the basis of association rules by comparing the antecedent part of rules with the known attributes values of missing value observation, (b) For the case, when there is no association rule exist or fired against any missing value (i.e. no relations exist between the attributes of other observations with the attributes of missing value), then the missing values are imputed using the imputation technique based on k-nearest neighbor approach [6, 7]. The main reason why we are using k-nearest approach as a hybrid combination is that, it is considered to be more robust against noise or in the case when the relationship between observations of dataset is very small [7]. Figure 1 shows the framework of missing values imputation using our hybrid imputation framework.

At the start of imputation process a set of strong (with good support and confidence) association rules are created on the basis of given training dataset with support and confidence threshold. Once association rules are created, the *HMiT* utilizes them for missing values imputation. For each observation *X* with missing values, association rules are fired by comparing the known attributes values of *X* with the antecedent part of association rules one by one. If the know attributes values of *X* are the subset of any association rule *R*, then *R* is added in the fired set *F*. Once all association rules are checked against the missing value of *X*, then the set *F* is considered for imputation. If the set *F* is non empty, then median and mod of the consequent part of the rules in set *F* is used for missing value imputation in case of numeric and discrete attributes. For the case, when the set $F$ is empty, then the missing value of $X$ is imputed using k-nearest neighbor approach. The pseudo code of our *HMiT* is described in Figure 2. Lines from 1 to 2 create the association rules on the basis of given support and confidence threshold. Lines from 4 to 7 first compare the antecedent part of all association rules with the known attributes values of missing value observation $X$. If the known attributes values of $X$ are subset of any rule $R$, then $R$ is added in the set $F$. If the set $F$ is non empty, then the missing values are impute on the basis of consequent part of fired association rules in Line 9 and 10, otherwise the missing values are impute using k-nearest neighbor approach in Line 12.

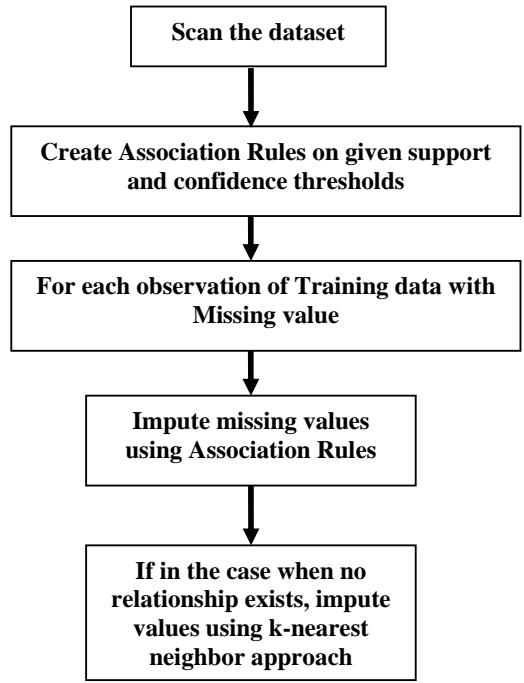

**Fig. 1.** The framework of our *HMiT* missing values imputation technique.

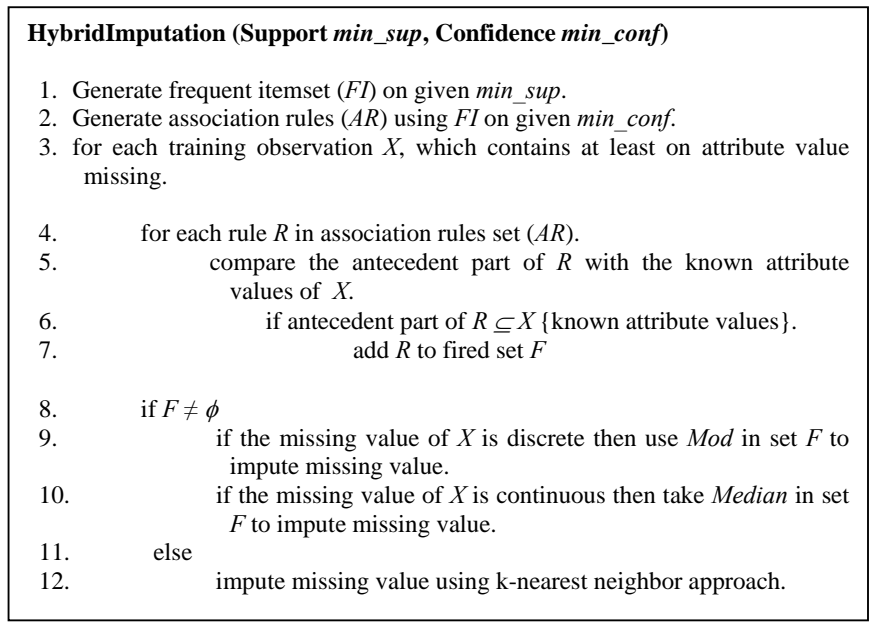

**Fig. 2.** Pseudo code of *HMiT* missing values imputation technique

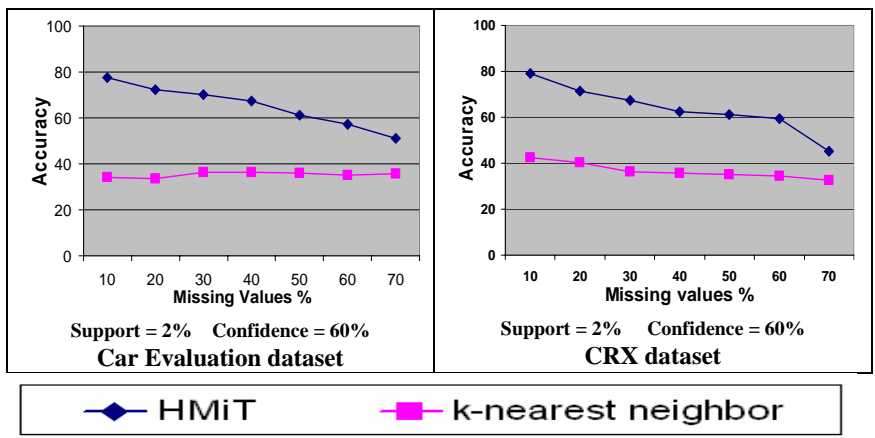

**Fig. 3.** Effect of random missing values on missing values imputation accuracy

## 3 Experimental Results

To evaluate the performance of *HMiT* we perform our experimental results on benchmark datasets available at [15]. The brief introduction of each dataset is described in Table 1. To validate the effectiveness of *HMiT*, we add random missing values in each of our experimental dataset. For performance reasons, we use Ramp [5] algorithm for frequent itemset mining and association rules generation. All the code of *HMiT* is written in Visual C++ 6.0, and the experiments are performed on 1.8 GHz machine with main memory of size 256 MB, running Windows XP 2005.

**Table 1.** Details of our experimental datasets

| Datasets | Instances | Total Attributes | Classes |
|---|---|---|---|
| Car Evaluation | 1728 | 6 | 4 |
| CRX | 690 | 16 | 2 |

### 3.1 Effect of Percentage of Missing Values on Missing Values Imputation Accuracy

The effect of percentage of missing values on imputation accuracy is shown in Figure 3 with *HMiT* and k-nearest neighbor approach. The results in Figure 3 are showing that as the level of percentage of missing values increases the accuracy of predicting missing values decreases. We obtain the Figure 3 results by fixing the support and confidence thresholds as 40 and 60% respectively and nearest neighbor size as 10. The reason behind why use nearest neighbor size = 10, is descried in [1]. From the results it is clear, that our *HMiT* generates good results as compared to k-nearest neighbor approach on all levels.

### 3.2 Effect of Confidence Threshold on Missing Values on Missing Values Imputation Accuracy

The effect of confidence threshold on missing values imputation accuracy is shown in the Figure 4. We obtain the Figure 4 results by fixing the support threshold as 40 and insert random missing values with 20%, while the confidence threshold was varied from 20% to 100%. For clear understanding, we exclude the accuracy effect of k-nearest neighbor from Figure 4 results. By looking the results, it is clear that as the confidence threshold increases the accuracy of predicting correctly missing also increases. For higher level of confidence only strong rules are generated and they more accurately predict the missing values, but in case of less confidence more weak or exceptional rules are generated which results in very less accuracy. From our experiments

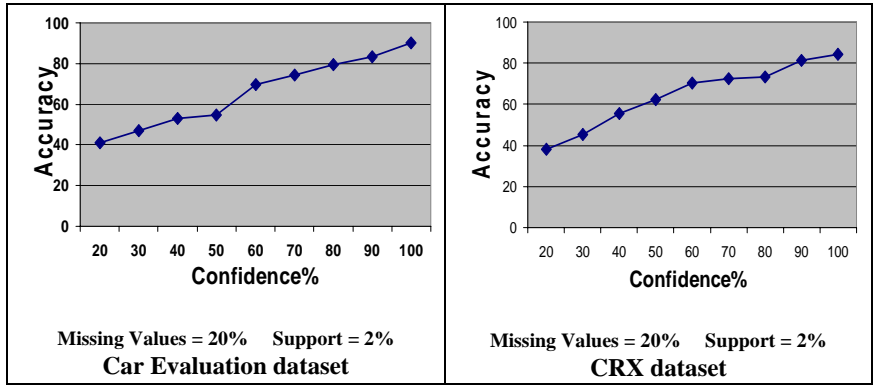

**Fig. 4.** Effect of confidence threshold on missing values imputation accuracy

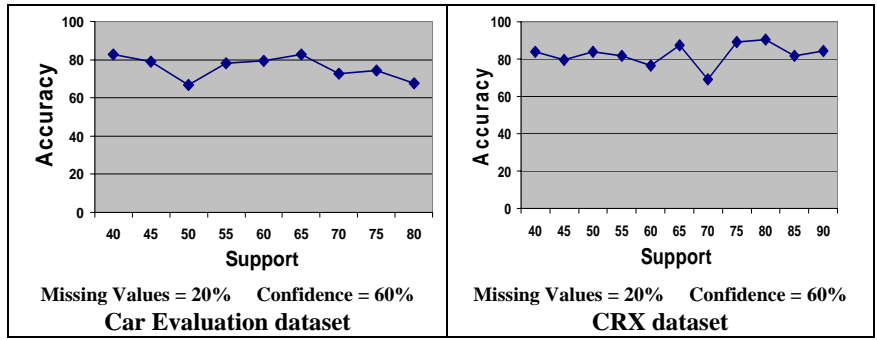

**Fig. 5.** Effect of support threshold on missing values imputation accuracy

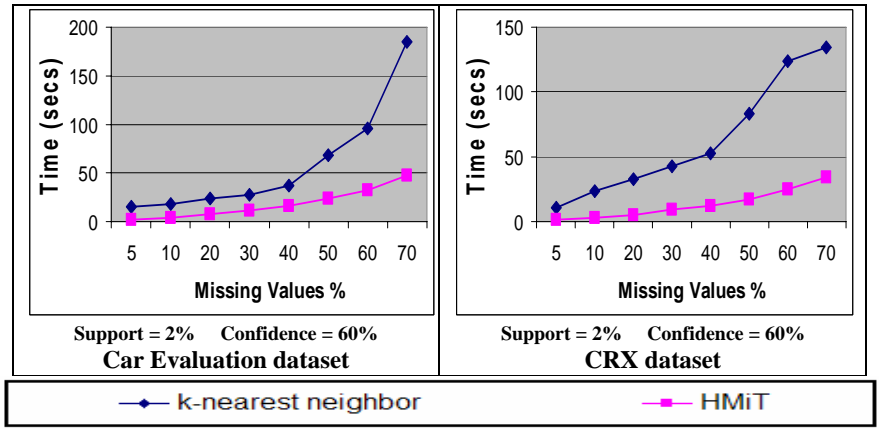

**Fig. 6.** Performance analysis of *HMiT* and k-nearest neighbor with different missing values level

we suggest that confidence threshold between 60% to 70% generates good results.

### 3.3 Effect of Support Threshold on Missing Values on Missing Values Imputation Accuracy

The effect of support threshold on missing values imputation accuracy is shown in the Figure 5. We obtain the Figure 5 results by fixing the confidence threshold as 60% and insert random missing values with 20%. Again for clear understanding, we exclude the accuracy effect of k-nearest neighbor from Figure 5 results. From the results it is clear that as the support threshold decreases more missing values are predicted with association rules, this is because more association rules are generated. But on the other hand as the support threshold increases less rules are generated and less number of missing values are predicted with association rules.

### 3.4 Performance Analysis of *HMiT* and k-nearest neighbor approach

The results of Figure 6 are showing that *HMiT* performs better in term of processing time as compared to k-nearest neighbor approach.

## 4 Conclusion

Missing value imputation is a complex problem in KDD and data mining tasks. In this paper we present a novel approach *HMiT* for missing values imputation based on association rule mining and hybrid combination of k-nearest neighbor approach. To analyze the effectiveness of *HMiT* we perform detail experiments results on benchmark datasets. Our results suggest that missing values imputation using our technique has good potential in term of accuracy and is also a good technique in term of processing time.